\documentclass{emulateapj}

\usepackage{natbib}

\newcommand{\DCO}{D$_{\rm CO}$}
\newcommand{\Hb}{H$\beta$}

\newcommand{\reduceme}{\mbox{R\raisebox{-0.35ex}{E}D%
\hspace{-0.05em}\raisebox{0.85ex}{uc}\hspace{-0.90em}%
\raisebox{-.35ex}{{m}}\hspace{0.05em}E}}

\shorttitle{Intermediate-age stellar populations in early-type galaxies}
\shortauthors{M{\'a}rmol-Queralt{\'o} et al.}

\begin{document}

\title{Evidence for intermediate-age stellar populations in early-type 
galaxies from K-band spectroscopy}

\author{E.~M{\'a}rmol-Queralt{\'o}\altaffilmark{1},
        N.~Cardiel\altaffilmark{1},
        P.~S{\'a}nchez-Bl{\'a}zquez\altaffilmark{2},
        S.C.~Trager\altaffilmark{3},
        R.F.~Peletier\altaffilmark{3},
        H.~Kuntschner\altaffilmark{4},
        D.R.~Silva\altaffilmark{5},
        A.J.~Cenarro\altaffilmark{2},
        A.~Vazdekis\altaffilmark{2},
        J.~Gorgas\altaffilmark{1}}
\altaffiltext{1}{Dpto. de Astrof\'{\i}sica, Universidad Complutense de Madrid, 
                 E28040, Madrid, Spain, \email{emq@astrax.fis.ucm.es}}
\altaffiltext{2}{Instituto de Astrof{\'{\i}}sica de Canarias. c/ V{\'{\i}}a
                 L{\'a}ctea S/N, E38205, La Laguna, Tenerife, Spain}
\altaffiltext{3}{Kapteyn Astronomical Institute, University of Groningen, 
                 Postbus 800,9700 Av Groningen, the Netherlands}
\altaffiltext{4}{Space Telescope -- European Coordinating Facility,
                 Karl-Schwarzschild-Str. 2, D-85748 Garching, Germany}
\altaffiltext{5}{National Optical Astronomy Observatory,
                 950 North Cherry Avenue, Tucson, AZ, 85719 USA}

\begin{abstract}

The study of stellar populations in early-type galaxies in different
environments is a powerful tool for constraining their star formation
histories. This study has been traditionally restricted to the optical range,
where dwarfs around the turn-off and stars at the base of the RGB dominate the
integrated light at all ages. The near-infrared spectral range is especially
interesting since in the presence of an intermediate-age population, AGB stars
are the main contributors. In this letter, we measure the near-infrared
indices NaI and \DCO\ for a sample of 12 early-type galaxies in low density
environments and compare them with the Fornax galaxy sample presented by
\citet{Fornax_red}. The analysis of these indices in combination with Lick/IDS
indices in the optical range reveals i) the NaI index is a metallicity
indicator as good as C4668 in the optical range, and ii) \DCO\ is a tracer of
intermediate-age stellar populations. We find that low-mass galaxies in low
density environments show higher NaI and \DCO\ than those located in Fornax
cluster, which points towards a late stage of star formation for the galaxies
in less dense environments, in agreement with results from other studies using
independent methods.

\end{abstract}


\section{Introduction}

For the majority of galaxies in the Universe, the analysis of the stellar
content is restricted to the integrated light of their stellar populations,
where a small percentage of luminous stars can change the spectra dramatically
and bias the final results. A way to overcome this bias is to use a
multiwavelength approach, since the fractional contribution from different
stars varies along the spectral range. However, most of the observational
effort has focused on the optical range, mainly due to the scarcity of
appropriate instrumentation in other spectral regions. Especially interesting
is the study of the near-infrared (near-IR) window, where red supergiants in
young populations ($< 1$~Gyr), asymptotic giant branch (AGB) stars in
intermediate-age populations ($\sim 1-2$~Gyr), and the tip of the red giant
branch (RGB) stars in the near-IR are the main contributors, as opposed
to turn-off dwarfs and stars at the lower part of the RGB in the optical.
Thus, near-IR spectroscopy may allow a breaking of the degeneracy between
multiple stellar populations that plagues optical absorption-line strength
studies \citep[e.g.,][]{Trager00b,SB06b}. In addition, the near-IR
wavelength range is less affected by line blanketing effects than other
spectral intervals.

\begin{figure}
\centering
\includegraphics[angle=-90,scale=0.80]{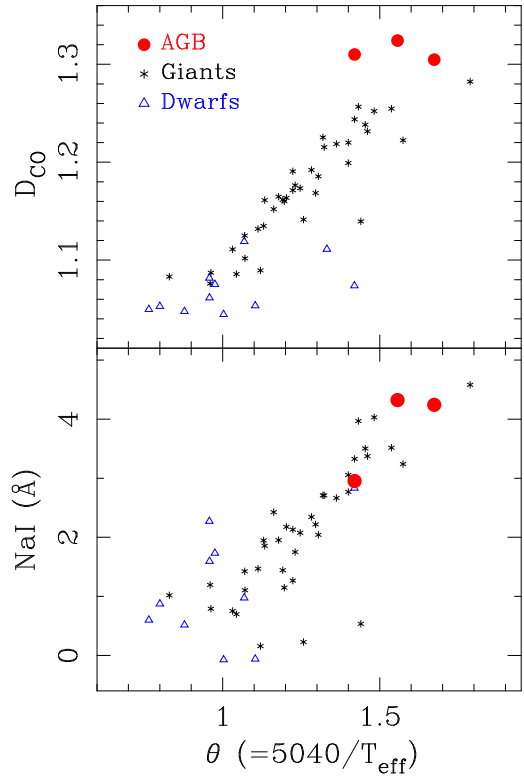} \caption{NaI and \DCO\
indices as a function of  $\theta = 5040/{\rm T_{eff}}$ for the subsample of
the K-band stellar library of MQ08  with the highest quality spectra. The
meaning of symbols and colors is explained in the inset. The AGB stars have
higher \DCO\ values than giants, while this behavior is not observed in
the NaI index.}
\label{fig1}
\end{figure}

\citet{libCO} (hereafter MQ08) presented a detailed study of the CO band at
2.3~$\mu$m as a function of the basic stellar atmospheric parameters (effective
temperature, surface gravity and metallicity) using a 
a new spectral library of stars in the K band. This work confirmed that
this CO absorption band is sensitive to metallicity and it also shows good
evidence for increased CO absorption strength in AGB stars compared to non-AGB
stars (see Figure~14 in MQ08 and Figure~\ref{fig1} in this letter). These
results confirm the suitability of the CO index for stellar population studies.
Furthemore, in this spectral range there is a prominent absorption feature at
2.2~$\mu$m mainly due to sodium and scandium \citep{WH96}. This feature is very
sensitive to effective temperature of the stars. Once that parameter is fixed,
AGB and RGB stars show very similar values (see Figure~\ref{fig1}).

The study of galaxies in different environments is a powerfull tool to
understand their evolution and star formation history \citep[e.g.,
S{\'a}nchez-Bl{\'a}zquez et al. 2003 --hereafer SB03--,][]{ESO_galaxies, Thomas05, SB06a, Collobert06}.
In this letter, we investigate possible differences in their stellar
populations comparing the more prominent absorption features in the K-band.
Only a few detailed studies have been presented in this spectral range after
the pioneering photometric work by \citet{Frogel78} and \citet{Aaronson78} in
early-type galaxies. The first detailed spectroscopic analysis in
the K-band of these galaxies depending on their environmnet was presented by
\citet{Mobasher96}, who used the strength of the first CO band at 2.3~$\mu$m as
an indicator of the presence of AGB stars and, therefore, of an
intermediate-age population in elliptical galaxies. They explored the influence
of environment finding, initially, a systematic difference between field and
cluster galaxies that was later contradicted in \citet{James99}, when the
authors enlarged their sample. They also studied the CO strength in Coma
cluster galaxies \citep{Mobasher00} finding variations in the strength of this
feature as a function of cluster-centric distance, with more distant galaxies
showing deeper CO absorption. 

\citet{Fornax_red} (hereafter S08) observed a sample of 11 early-type galaxies
(see Table~\ref{table}) in the Fornax cluster, and analyzed several
line-strength indices (NaI, CaI, $\langle \mbox{FeI} \rangle$ and CO) in the K
band. For those galaxies showing no trace of young or intermediate-age
populations, they found a strong correlation between NaI and CO with velocity
dispersion ($\sigma$), as well as between those indices and the optical
metallicity indicator [MgFe]$'$ \citep{Thomas03}. However, the galaxies
with young populations in their sample separate from the global trends. More
recently, \citet{Cesetti09} obtained low resolution espectra for 14 early-type
galaxies and confirmed the relation between the near-IR and the optical
features.

To unravel possible differences in the stellar content of galaxies in different
environments, we analyze a sample of elliptical galaxies located in low density
environments (hereafter {\it field} galaxies\footnote{This sample includes
two galaxies in the Virgo cluster. SB03 showed that there is no differences in
their optical indices when comparing with more isolated galaxies, and for that
reason they have been included here.}) in comparison with the Fornax galaxy
sample presented by S08. This letter presents the first results of this
spectroscopic study, focusing on the NaI and the CO indices. The rest of the
indices, as well as a detailed description of the data reduction, will be
presented in a forthcoming paper (M\'{a}rmol-Queralt\'{o} et al., 2009, in
preparation). In \S~2 we describe the data used in this study, while the
results and discussion are presented in \S~3 and \S~4, respectively. Finally,
in \S~5 we present the main conclusions derived from this work.

\section{The data}

Long-slit spectra for 12 early-type field galaxies (see Table~\ref{table}) from
the samples of \citet{SB06a} and \citet{ESO_galaxies} plus 2 Fornax galaxies
from S08 (for comparison purposes) were obtained with the ISAAC near-IR imaging
spectrometer mounted at 8.2~m UT1/Antu telescope at Cerro de Paranal (Chile)
with exactly the same instrumental configuration employed by S08. Thus, both
samples can be readily compared. The field galaxy sample was chosen to cover a
range in $\sigma$ similar to that spanned by S08's Fornax sample. The
spectrometer was used in medium-resolution mode (SWS1 -- MR), providing a
resolution of 7.1~\AA\, at $2.3\,\mu$m (FWHM) with a reciprocal dispersion of
1.21~\AA/pixel for a $120 \arcsec \times 1 \arcsec$ slit. The central
wavelength was chosen for each galaxy to include both Na{\sc i} features and
the first CO band ($\sim 2.20\,\mu$m and $\sim 2.29\,\mu$m at restframe,
respectively). For each object, several exposures along the slit (A and B
positions) were taken for sky substraction, as is the usual practice for
infrared observations. The slit was oriented along the minor axis of the
galaxies when possible. After each galaxy set, a nearby A-type star was
observed for relative flux calibration and telluric correction. Finally,
halogen lamps (on and off) and arc lamps were observed for flat-fielding, and
C-distortion correction and wavelength calibration, respectively.

We carried out a standard data reduction in the infrared using \reduceme\
\citep{Cardiel99}. The details of the process will be presented in a
forthcoming paper. Galaxy spectra were extracted within a radius corresponding
to $1/8 R_{\mbox{\scriptsize eff}}$ at the observed position angle for
each galaxy, where $R_{\mbox{\scriptsize eff}}$ is the effective radius.

We measured the NaI index \citep{Frogel2001} and the \DCO\ index (MQ08). The
NaI measurements were corrected for broadening effects using the
$\sigma$-correction computed by S08, while the \DCO\ index is almost
insensitive to this effect (see MQ08). In Table~\ref{table} we list these
measurements for all the galaxies, together with the central velocity
dispersion ($\sigma$) computed for each galaxy from these infrared spectra.
Previous optical work is available for all galaxies. The optical Lick/IDS
indices Mgb, C4668 and \Hb\ were measured on the spectra extracted within a
radius corresponding to $1/8 R_{\mbox{\scriptsize eff}}$ in the original
data by \citet{SB03} \citep[see also][]{SB06a} for the field galaxies. For
Fornax cluster galaxies, Mgb and \Hb\ were taken from S08 (measured on the
spectra extracted within $1/8 R_{\mbox{\scriptsize eff}}$) and C4668
from \citet{Fornax_optico} (original data). All these optical indices have also
been included in Table~\ref{table}. 

\section{Results}

\begin{figure}
\centering
\includegraphics[angle=-90,scale=1.30]{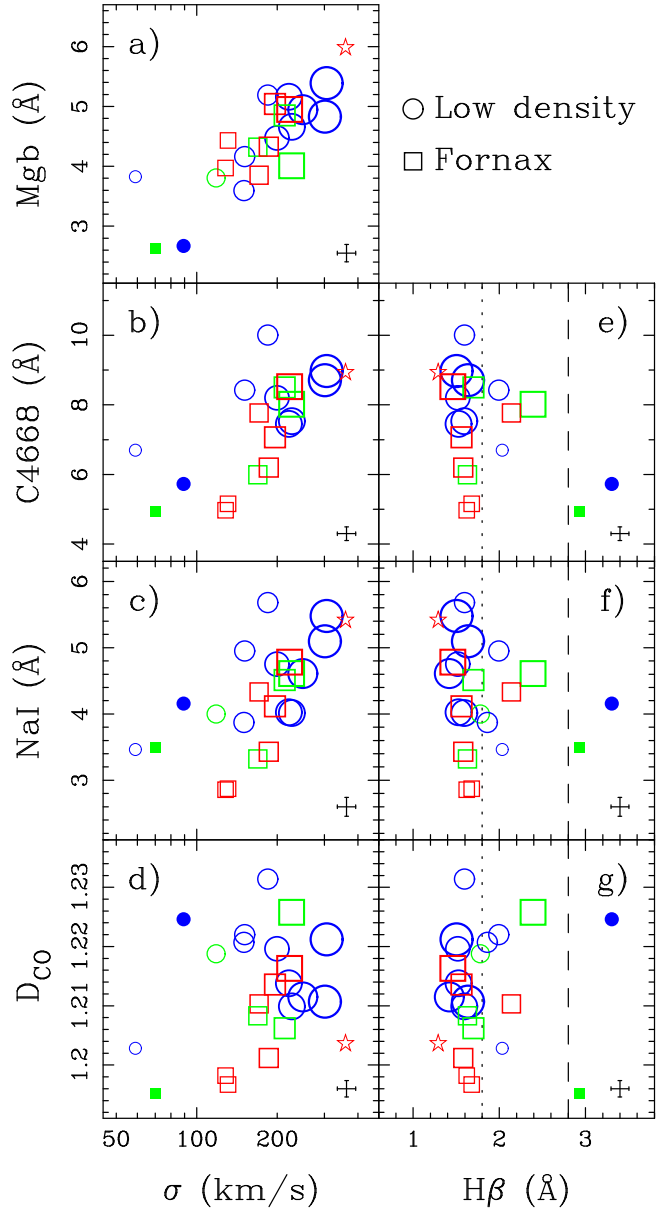}
\caption{{\it Left column}: Line-strength indices versus $\sigma$ for the
galaxies analyzed in this letter (data in Table~\ref{table}). {\it Right
column}: Line-strength indices versus \Hb\ index. The dashed line indicates the
limit for considering very young galaxies in the blue spectral range. The
dotted line shows the limit for H$\beta > 1.8$~\AA. Blue circles correspond to
galaxies in the field, while red squares represent Fornax cluster
galaxies.  S0 galaxies of both samples are plotted in green. The two galaxies
with the highest H$\beta$ values are marked with filled symbols. The star
represents the cD galaxy NGC1399 \citep[see][for an extensive study of this
object in the infrared]{Lyubenova08}. Symbol size increases according to the
central velocity dispersion of the galaxies.}
\label{fig2}
\end{figure}

Since there are no reliable stellar population models available describing the
behaviour of the near-IR indices used in this study, we will interpret the
measured features by comparing them with other, better understood, indices. Due
to the very strong correlation between metallicity and $\sigma$ for Fornax
elliptical galaxies \citep{Fornax_optico}, we use $\sigma$ as a ``proxy'' for
global metallicity in these galaxies. As a ``proxy'' for the mean age, we will
use the \Hb\ index, which is an indicator of the temperature of the turn-off
and, therefore, mainly sensitive to the presence of young stellar populations
\citep[between $\sim 2\times10^8$ and $10^9$ yr -- e.g.][]{Allard06}.
Figure~\ref{fig2} shows the behaviour of different indices against $\sigma$
(left column) and \Hb\ (right column). We note that the number of galaxies
analyzed here is small and, therefore, our results might not necessarily
apply to all early-type galaxies

There is a clear correlation between the indices Mgb, C4668, NaI, \DCO, and
$\sigma$ for the Fornax galaxies (already shown by \cite{Fornax_optico} for the
optical indices and S08 for the near-IR features). For the field sample, the
indices Mgb, C4668 and NaI are also correlated with $\sigma$, and these
relations exhibit a larger scatter than the ones obtained for the Fornax sample
(see a quantitative statistical analysis of all these diagrams in
Table~\ref{statis}). We do not find differences in the Mgb measurements
between both galaxy samples, while field galaxies exhibit higher C4668, NaI and
\DCO\ indices than Fornax galaxies at \mbox{$\sigma < 200$~km~s$^{-1}$}, with
the largest differences for the CO feature. This behaviour indicates that there
are differences in the star formation history and the metal enrichment
between both galaxy samples. As we will discuss below, we are likely seeing the
effect of more metal rich intermediate-age stellar populations in these field
galaxies, which are not present in most of their Fornax counterparts (S08). For
galaxies with higher $\sigma$, both samples seem to follow the same trend.

In the right column of Figure~\ref{fig2} we compare different indices with \Hb.
We do not find any correlation. Two of the galaxies --marked in
all the plots with filled symbols-- show very high \Hb\ indices. These galaxies
deviate from all the relations defined by the indices and other parameters. In
addition, there are several galaxies with $1.8 < {\rm H}\beta < 2.8$~\AA, while
the bulk of the galaxies have lower \Hb\ indices. S08 already showed the
presence of a small amount of intermediate-age population in the three Fornax
galaxies with \Hb~$>1.8$~\AA.

\begin{figure}
\centering
\includegraphics[angle=-90,scale=1.30]{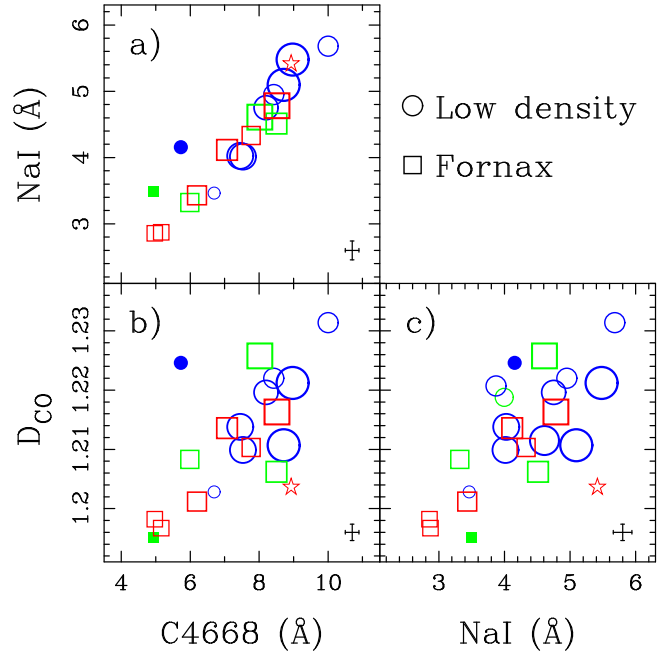}
\caption{Index-index diagrams for the galaxies in this study. Symbols are
explained in Figure~\ref{fig1}.}
\label{fig3}
\end{figure}

Since we find differences in C4668, NaI and \DCO\ between galaxies in low
density environment and in Fornax cluster, we explore possible dependences
among these indices in Figure~\ref{fig3}. The most interesting result is the
extremely tight correlation between NaI and C4668 indices in Figure~\ref{fig3}a.
This is remarkable given the fact that the indices have been measured on
completely independent spectra processed by different authors in distinct
spectral ranges. Interestingly, the two galaxies with highest \Hb\ separate
from the global trend. Finally, there are also correlations between \DCO\ and
the indices C4668 and NaI (Figs.~\ref{fig3}b and~\ref{fig3}c) (see 
Table~\ref{statis}). 

\section{Discussion}

The main two results of this work are (1) the excellent correlation
between the NaI and C4668 indices, and (2) the higher \DCO\ and NaI values in
the low-sigma field galaxies compared with their Fornax counterparts.  How can
we interpret these results in terms of physical parameters? To answer
this question, we need to take into account the known dependences of the
indices to several parameters. We have to warn the reader that this section is
rather speculative, because of the lack of stellar population models in this
wavelength region. We should also note that a simple discussion in terms of
SSPs ({\it Simple Stellar Populations}) is inadequate here, since, as we will
discuss, the measured indices in Fornax and field galaxies are very likely
indicating differences in their star formation history.

First we describe the qualitative behaviour of the indices. For an old SSP we
deduce, from the work by MQ08, that \DCO\ increases as a function of
metallicity. For younger stellar populations, \DCO\ increases due to the
presence of AGB stars. The NaI index increases with metallicity for old stellar
populations (see S08). In S08 it was also shown that NaI increases when a
young population is present in the galaxy (see Figure 16 in that paper). Mgb
and C4668, like the rest of the non-Balmer indices in the Lick system, are
weaker in younger stellar populations \citep[e.g.][]{Vazdekis03, Bruzual03,
Thomas03, Schiavon07}. These indices are also sensitive to metallicity,
with higher values for more metal-rich populations. 

How to explain the behaviour of the line indices shown in Figure~\ref{fig2}?
Following previous studies analyzing galaxies in different environments
\citep[e.g., SB03][]{Trager00b, ESO_galaxies, Thomas05, SB06a, SB06b,
Collobert06} we start with the assumption that the differences between the
field and Fornax samples could be due to distinct star formation histories. As
\citet{Fornax_optico} showed for the Fornax cluster, elliptical galaxies and
lenticulars of high $\sigma$ form an almost coeval family of galaxies, and they
exhibit a wide range in metallicity which explains the good correlation of all
the indices with $\sigma$. On the other hand, previous works have
suggested an extended star formation history in field galaxies
\citep[e.g.,SB03][]{Collobert06, Toloba09}. If this is the case, we would
expect two effects: i) the chemical enrichment of the interstellar medium
produced by previous generations of stars. As a result, the new generation of
stars will be more metal rich than the previous ones. ii) The presence of AGB
stars from the intermediate-age population of recent star formation episodes.

The higher \DCO\ values for field galaxies with \mbox{$\sigma <
200$~km~s$^{-1}$} can be explained by a late stage star formation in these
galaxies, which is reflected in a more metal rich population and the presence
of AGB stars. Since this index has a positive response to both effects, the
differences between field and Fornax galaxies are more prominent here than in
other indices. In addition, the more metal rich population in field galaxies
would be responsible for the high values of NaI and C4668 in these objects (see
Figure~\ref{fig2}b-c and Figure~\ref{fig3}a-c), since both indices are very
sensitive to metallicity. Actually, the excellent correlation between NaI
and C4668 shown in Figure~\ref{fig3}a might be indicating that the effect of an
intermediate-age population in these two indices is less important, and,
therefore, they are both excellent metallicity indicators for old and
intermediate ages. However, the two galaxies with the highest \Hb\ separate from
the global trend, as expected due to the sensitivity of these indices to the
presence of very young and more metal rich stars, as explained above. The
larger scatter in \DCO\ vs.\ C4668 diagram (Figure~\ref{fig3}b) can be 
explained by considering that \DCO\ is sensitive to the presence of AGB stars,
contrary to the optical C4668 index. Last, both galaxy samples exhibit a
similar Mgb index. This is explained by the age-metallicy degeneracy described
in previous works \citep[e.g.][]{Worthey92, Pedraz98, Jorgensen99,
Kuntschner01}. 

\section{Conclusions}

This work has shown the crucial role that the near-IR line-strength
indices can play in the understanding of the star formation histories of
early-type galaxies. Although interpreting these near-IR indices is difficult
without models and the analyzed galaxy samples are still small, we have
demonstrated that the measurement of line-strength indices in the near-IR is a
useful tool to constrain the star formation histories of these galaxies. In
particular, the NaI index has been shown to be a metallicity indicator as good
as C4668 in the optical range, while the \DCO\ index can be used as tracer of
intermediate-age stellar populations. The differences in C4668, NaI, and \DCO, 
when studying low-mass galaxies in
Fornax cluster and in low density environments, can be interpreted as field
galaxies having undergone later stellar formation episodes than Fornax
galaxies. Once synthesis stellar population models in the near-IR range are
available, more detailed and quantitative predictions will be made using this
interesting wavelength region.

\acknowledgments

The authors acknowledge the referee, Bahram Mobasher, his useful comments. EMQ
acknowledges the SMES for a FPI PhD fellowship. This work has been partially
supported by the SMES through the research project AYA2007-67752-C03-01. NC
also acknowledges financial support from the research project
AYA2006-15698-C02-02. PSB acknowledges the finantial support from a Marie Curie
European Reintegration Grant within the 7th European Community Framework
Programme.\\ 

{\it Facilities:} \facility{VLT:Antu (ISAAC)}



\begin{deluxetable}{lccccccccc}
\tabletypesize{\scriptsize}
\tablecaption{Measurements for the early-type galaxies in this study}
\tablehead{
  \colhead{Galaxy} &  
  \colhead{Type} &  
  \colhead{\Hb\tablenotemark{a}} & 
  \colhead{Mgb\tablenotemark{a}}  &  
  \colhead{C4668\tablenotemark{b}} &  
  \colhead{NaI} &  
  \colhead{\DCO} &
  \colhead{$\sigma$\tablenotemark{d}} &
  \colhead{Environment\tablenotemark{e}} \\
  \colhead{} &  
  \colhead{} &  
  \colhead{(\AA)} & 
  \colhead{(\AA)} &  
  \colhead{(\AA)} &  
  \colhead{(\AA)} &  
  \colhead{} &
  \colhead{(km~s$^{-1}$)} &
  \colhead{} 
 }
\startdata
NGC3605    & E4-5 & $ 2.03 \pm 0.10$ & $ 3.83 \pm 0.15$ & $ 6.70 \pm 0.30$ & $ 3.46 \pm 0.08$ & $1.2028 \pm 0.0016$ & $ 59.1 \pm  7.0$  & Low density \\
NGC3818    &  E5  & $ 1.52 \pm 0.05$ & $ 4.47 \pm 0.09$ & $ 8.20 \pm 0.21$ & $ 4.75 \pm 0.12$ & $1.2196 \pm 0.0025$ & $199.5 \pm 20.7$  & Low density \\
NGC4261    & E2-3 & $ 1.63 \pm 0.05$ & $ 4.83 \pm 0.12$ & $ 8.71 \pm 0.21$ & $ 5.10 \pm 0.26$ & $1.2107 \pm 0.0022$ & $300.9 \pm 29.4$  & Low density \\
NGC4564    & E6   & $ 1.60 \pm 0.05$ & $ 5.19 \pm 0.11$ & $10.01 \pm 0.20$ & $ 5.68 \pm 0.14$ & $1.2314 \pm 0.0019$ & $184.3 \pm 24.8$  & Low density \\
NGC4636    & E-S0 & $ 1.53 \pm 0.09$ & $ 5.16 \pm 0.15$ & $ 7.45 \pm 0.41$ & $ 4.03 \pm 0.14$ & $1.2138 \pm 0.0019$ & $221.2 \pm 23.7$  & Low density \\
NGC4742    & E4   & $ 3.30 \pm 0.11$ & $ 2.67 \pm 0.11$ & $ 5.72 \pm 0.20$ & $ 4.16 \pm 0.03$ & $1.2246 \pm 0.0009$ & $ 89.2 \pm  9.6$  & Low density \\
NGC5796    & E    & $ 1.50 \pm 0.07$ & $ 5.39 \pm 0.16$ & $ 8.97 \pm 0.23$ & $ 5.48 \pm 0.21$ & $1.2212 \pm 0.0021$ & $305.4 \pm 22.3$  & Low density \\
NGC5813    & E1-2 & $ 1.59 \pm 0.05$ & $ 4.66 \pm 0.12$ & $ 7.53 \pm 0.21$ & $ 4.01 \pm 0.14$ & $1.2099 \pm 0.0021$ & $226.7 \pm 25.7$  & Low density \\
NGC5831    & E3   & $ 1.99 \pm 0.15$ & $ 4.16 \pm 0.12$ & $ 8.43 \pm 0.22$ & $ 4.95 \pm 0.08$ & $1.2220 \pm 0.0019$ & $151.0 \pm 14.5$  & Low density \\
ESO382-G16 & E    & $ 1.42 \pm 0.11$ & $ 4.94 \pm 0.15$ &       \dots      & $ 4.61 \pm 0.14$ & $1.2115 \pm 0.0021$ & $249.2 \pm 21.0$  & Low density \\
ESO446-G49 & S0   & $ 1.78 \pm 0.12$ & $ 3.81 \pm 0.14$ &       \dots      & $ 4.00 \pm 0.11$ & $1.2188 \pm 0.0027$ & $118.2 \pm 12.6$  & Low density \\
ESO503-G12 & E    & $ 1.86 \pm 0.10$ & $ 3.59 \pm 0.13$ &       \dots      & $ 3.87 \pm 0.08$ & $1.2207 \pm 0.0016$ & $150.1 \pm 17.9$  & Low density \\
NGC1316    & S0pec& $ 2.39 \pm 0.11$ & $ 4.01 \pm 0.13$ & $ 8.02 \pm 0.19$ & $ 4.61 \pm 0.15$ & $1.2258 \pm 0.0015$ & $226.0 \pm  8.0$  &  Fornax \\
NGC1344    & E5  & $ 2.14 \pm 0.09$ & $ 3.85 \pm 0.10$ &        \dots     & $ 4.33 \pm 0.15$ & $1.2103 \pm 0.0005$ & $171.0 \pm  8.0$   & Fornax \\
NGC1374    & E0  & $ 1.56 \pm 0.15$ & $ 5.04 \pm 0.20$ & $ 7.07 \pm 0.24$ & $ 4.12 \pm 0.13$ & $1.2136 \pm 0.0006$ & $196.0 \pm  9.0$   & Fornax \\
NGC1375    & S0  & $ 2.93 \pm 0.12$ & $ 2.62 \pm 0.16$ & $ 4.93 \pm 0.22$ & $ 3.50 \pm 0.23$ & $1.1952 \pm 0.0006$ & $ 70.0 \pm  9.0$   & Fornax \\
NGC1379\tablenotemark{$\star$} & E0 & $ 1.68 \pm 0.12$ & $ 4.43 \pm 0.15$ & $ 5.16 \pm 0.23$ & $ 2.87 \pm 0.19$ & $1.1967 \pm 0.0006$ & $130.8 \pm  8.2$   & Fornax \\
NGC1380    & S0a & $ 1.70 \pm 0.15$ & $ 4.85 \pm 0.21$ & $ 8.50 \pm 0.29$ & $ 4.51 \pm 0.10$ & $1.2062 \pm 0.0002$ & $213.0 \pm  6.0$   & Fornax \\
NGC1381    & S0  & $ 1.63 \pm 0.09$ & $ 4.32 \pm 0.12$ & $ 5.99 \pm 0.16$ & $ 3.32 \pm 0.15$ & $1.2083 \pm 0.0005$ & $169.0 \pm  8.0$   & Fornax \\
NGC1399    & E0  & $ 1.29 \pm 0.16$ & $ 5.98 \pm 0.24$ & $ 8.93 \pm 0.39$ & $ 5.42 \pm 0.23$ & $1.2036 \pm 0.0022$ & $360.0 \pm  8.0$   & Fornax \\
NGC1404\tablenotemark{$\star$} & E2 & $ 1.46 \pm 0.11$ & $ 4.95 \pm 0.15$ & $ 8.52 \pm 0.24$ & $ 4.78 \pm 0.08$ & $1.2163 \pm 0.0006$ & $222.2 \pm  7.7$   & Fornax \\
NGC1419    & E0  & $ 1.62 \pm 0.18$ & $ 3.97 \pm 0.23$ & $ 4.97 \pm 0.24$ & $ 2.86 \pm 0.17$ & $1.1982 \pm 0.0004$ & $128.0 \pm  6.0$   & Fornax \\
NGC1427    & E4  & $ 1.58 \pm 0.08$ & $ 4.33 \pm 0.11$ & $ 6.20 \pm 0.14$ & $ 3.43 \pm 0.17$ & $1.2012 \pm 0.0004$ & $186.0 \pm 11.0$   & Fornax \\
\enddata
\tablenotetext{a}{\Hb\ and Mgb indices measured on the optical spectra,
extracted within $1/8 R_{\mbox{\scriptsize eff}}$, by SB03 and
\citet{Fornax_optico}.}
\tablenotetext{b}{C4668 index. New measurements on the optical spectra
extracted within $1/8 R_{\mbox{\scriptsize eff}}$ by SB03 and original data
for Fornax galaxies \citep{Fornax_optico}.}
\tablenotetext{d}{Central velocity dispersion derived from the near-infrared
data by S08 and M\'{a}rmol-Queralt\'{o} et al.\ (2009, in preparation).}
\tablenotetext{e}{Following the criteria in \citet{SB06a} and
\citet{ESO_galaxies}.}
\tablenotetext{$\star$}{Fornax galaxies also observed in this study.}
\label{table}
\end{deluxetable}

\begin{deluxetable}{l@{$\;\;$}c@{$\;\;$}c@{$\;\;$}cc@{$\;\;$}c@{$\;\;$}c}
\tabletypesize{\scriptsize}
\tablecaption{Statistical analysis of the correlations for field and Fornax
galaxies}
\tablehead{
  \colhead{} &  
  \colhead{} &  
  \colhead{\makebox[0pt][c]{Low density}} &  
  \colhead{} & 
  \colhead{} &  
  \colhead{\makebox[0pt][c]{Fornax}} &  
  \colhead{} \\   
  \colhead{Diagram} &  
  \colhead{$r_S$} & 
  \colhead{$\alpha$}  &  
  \colhead{s$_{\rm r}$} &  
  \colhead{$r_S$} &  
  \colhead{$\alpha$} &  
  \colhead{s$_{\rm r}$}    
}
\startdata
Figure~2a: Mgb--$\sigma$   &   0.7636 &  0.0031&  0.6685&  0.5636&  0.0449&  0.4764\\
Figure~2b: C4668--$\sigma$ &   0.6455 &  0.0160&  1.0440&  0.9273& 5E$-05$&  0.3287\\
Figure~2c: NaI--$\sigma$   &   0.6091 &  0.0233&  1.4256&  0.9515& 1E$-05$&  0.2165\\
Figure~2d: \DCO--$\sigma$  &  \makebox[0pt][r]{$-$}0.0545$^\dag$ &  0.4367&  2.7721&  0.5757&  0.0408 &  1.5668\\
\hline                     
Figure~3a: NaI--C4668      &   0.9818 & 4E$-08$&  0.0531&  0.9879& 5E$-08$&  0.0787\\
Figure~3b: \DCO--C4668     &   0.5182 &  0.0512&  0.8005&  0.5394&  0.0538&  1.6243\\
Figure~3c: \DCO--NaI       &   0.5727 &  0.0328&  1.0957&  0.6000&  0.0333&  1.9062\\
\enddata
\tablenotetext{}{\mbox{}\hspace{-2mm}\parbox{97mm}{%
For each diagram, indicated in the first column of this table, the Spearman rank
correlation coefficient $r_S$ and the associated significance level $\alpha$ are
given, separately, for the field and Fornax subsamples. In all the cases, the
two galaxies with higher \Hb\ values (filled symbols in the diagrams) have been
excluded. With the aim of comparing the scatter observed in each galaxy
subsample the residual standard deviation $s_{\rm r}$ has also been determined
as the mean quadratic distance from a linear fit to the data points. Since the
data have uncertainties in both axes, the fits have been computed using an
ordinary least squares (OLS) bisector method \citep{Isobe90}. The data ranges
were previously renormalized for the proper measure of the perpendicular
distances from the data to the fitted straight lines. This renormalization
prevents the different weighting introduced by the absolute values of the data.
Notice that, despite the relations between the indices are evident from the
figures, some correlations are not significant due to size of the samples. 
$^\dag$ If the field galaxy with lowest $\sigma$, NGC3605, is also excluded, we
obtain $r_{\rm S} = -0.4061$ ($\alpha = 0.1221$).  Regarding its position in the
Mgb-- and C4668$-\sigma$ diagrams, this galaxy could belong to the new family of
high-metallicity low-luminosity galaxies found by \citet{Sansom08}. 
}
}
\label{statis}
\end{deluxetable}


\begin{thebibliography}{0}
\expandafter\ifx\csname natexlab\endcsname\relax\def\natexlab#1{#1}\fi
\expandafter\ifx\csname bibnamefont\endcsname\relax
  \def\bibnamefont#1{#1}\fi
\expandafter\ifx\csname bibfnamefont\endcsname\relax
  \def\bibfnamefont#1{#1}\fi
\expandafter\ifx\csname citenamefont\endcsname\relax
  \def\citenamefont#1{#1}\fi
\expandafter\ifx\csname url\endcsname\relax
  \def\url#1{\texttt{#1}}\fi
\expandafter\ifx\csname urlprefix\endcsname\relax\def\urlprefix{URL }\fi
\providecommand{\bibinfo}[2]{#2}
\providecommand{\eprint}[2][]{\url{#2}}

\end{thebibliography}


\begin{thebibliography}{32}
\expandafter\ifx\csname natexlab\endcsname\relax\def\natexlab#1{#1}\fi

\bibitem[{{Aaronson} {et~al.}(1978){Aaronson}, {Cohen}, {Mould}, \&
  {Malkan}}]{Aaronson78}
{Aaronson}, M., {Cohen}, J.~G., {Mould}, J., \& {Malkan}, M. 1978, \apj, 223,
  824

\bibitem[{{Allard} {et~al.}(2006){Allard}, {Knapen}, {Peletier}, \&
  {Sarzi}}]{Allard06}
{Allard}, E.~L., {Knapen}, J.~H., {Peletier}, R.~F., \& {Sarzi}, M. 2006,
  \mnras, 371, 1087

\bibitem[{{Bruzual} \& {Charlot}(2003)}]{Bruzual03}
{Bruzual}, G. \& {Charlot}, S. 2003, \mnras, 344, 1000

\bibitem[{{Cardiel}(1999)}]{Cardiel99}
{Cardiel}, N. 1999, Ph.D.~Thesis

\bibitem[{{Cesetti} {et~al.}(2009){Cesetti}, {Ivanov}, {Morelli}, {Pizzella},
  {Buson}, {Corsini}, {Dalla Bont{\`a}}, {Stiavelli}, \& {Bertola}}]{Cesetti09}
{Cesetti}, M., {Ivanov}, V.~D., {Morelli}, L., {Pizzella}, A., {Buson}, L.,
  {Corsini}, E.~M., {Dalla Bont{\`a}}, E., {Stiavelli}, M., \& {Bertola}, F.
  2009, \aap, 497, 41

\bibitem[{{Collobert} {et~al.}(2006){Collobert}, {Sarzi}, {Davies},
  {Kuntschner}, \& {Colless}}]{Collobert06}
{Collobert}, M., {Sarzi}, M., {Davies}, R.~L., {Kuntschner}, H., \& {Colless},
  M. 2006, \mnras, 370, 1213

\bibitem[{{Frogel} {et~al.}(1978){Frogel}, {Persson}, {Matthews}, \&
  {Aaronson}}]{Frogel78}
{Frogel}, J.~A., {Persson}, S.~E., {Matthews}, K., \& {Aaronson}, M. 1978,
  \apj, 220, 75

\bibitem[{{Frogel} {et~al.}(2001){Frogel}, {Stephens}, {Ram{\'{\i}}rez}, \&
  {DePoy}}]{Frogel2001}
{Frogel}, J.~A., {Stephens}, A., {Ram{\'{\i}}rez}, S., \& {DePoy}, D.~L. 2001,
  \aj, 122, 1896

\bibitem[{{Isobe} {et~al.}(1990){Isobe}, {Feigelson}, {Akritas}, \&
  {Babu}}]{Isobe90}
{Isobe}, T., {Feigelson}, E.~D., {Akritas}, M.~G., \& {Babu}, G.~J. 1990, \apj,
  364, 104

\bibitem[{{James} \& {Mobasher}(1999)}]{James99}
{James}, P.~A. \& {Mobasher}, B. 1999, \mnras, 306, 199

\bibitem[{{J{\o}rgensen} {et~al.}(1999){J{\o}rgensen}, {Franx}, {Hjorth}, \&
  {van Dokkum}}]{Jorgensen99}
{J{\o}rgensen}, I., {Franx}, M., {Hjorth}, J., \& {van Dokkum}, P.~G. 1999,
  \mnras, 308, 833

\bibitem[{{Kuntschner}(2000)}]{Fornax_optico}
{Kuntschner}, H. 2000, \mnras, 315, 184

\bibitem[{{Kuntschner} {et~al.}(2001){Kuntschner}, {Lucey}, {Smith}, {Hudson},
  \& {Davies}}]{Kuntschner01}
{Kuntschner}, H., {Lucey}, J.~R., {Smith}, R.~J., {Hudson}, M.~J., \& {Davies},
  R.~L. 2001, \mnras, 323, 615

\bibitem[{{Kuntschner} {et~al.}(2002){Kuntschner}, {Smith}, {Colless},
  {Davies}, {Kaldare}, \& {Vazdekis}}]{ESO_galaxies}
{Kuntschner}, H., {Smith}, R.~J., {Colless}, M., {Davies}, R.~L., {Kaldare},
  R., \& {Vazdekis}, A. 2002, \mnras, 337, 172

\bibitem[{{Lyubenova} {et~al.}(2008){Lyubenova}, {Kuntschner}, \&
  {Silva}}]{Lyubenova08}
{Lyubenova}, M., {Kuntschner}, H., \& {Silva}, D.~R. 2008, \aap, 485, 425

\bibitem[{{M{\'a}rmol-Queralt{\'o}} {et~al.}(2008){M{\'a}rmol-Queralt{\'o}},
  {Cardiel}, {Cenarro}, {Vazdekis}, {Gorgas}, {Pedraz}, {Peletier}, \&
  {S{\'a}nchez-Bl{\'a}zquez}}]{libCO}
{M{\'a}rmol-Queralt{\'o}}, E., {Cardiel}, N., {Cenarro}, A.~J., {Vazdekis}, A.,
  {Gorgas}, J., {Pedraz}, S., {Peletier}, R.~F., \& {S{\'a}nchez-Bl{\'a}zquez},
  P. 2008, \aap, 489, 885 (MQ08)

\bibitem[{{Mobasher} \& {James}(1996)}]{Mobasher96}
{Mobasher}, B. \& {James}, P.~A. 1996, \mnras, 280, 895

\bibitem[{{Mobasher} \& {James}(2000)}]{Mobasher00}
---. 2000, \mnras, 316, 507

\bibitem[{{Pedraz} {et~al.}(1998){Pedraz}, {Gorgas}, {Cardiel}, \&
  {Guzm{\'a}n}}]{Pedraz98}
{Pedraz}, S., {Gorgas}, J., {Cardiel}, N., \& {Guzm{\'a}n}, R. 1998, \apss,
  263, 159

\bibitem[{{S{\' a}nchez-Bl{\' a}zquez} {et~al.}(2003){S{\' a}nchez-Bl{\'
  a}zquez}, {Gorgas}, {Cardiel}, {Cenarro}, \& {Gonz{\' a}lez}}]{SB03}
{S{\' a}nchez-Bl{\' a}zquez}, P., {Gorgas}, J., {Cardiel}, N., {Cenarro}, J.,
  \& {Gonz{\' a}lez}, J.~J. 2003, \apjl, 590, L91 (SB03)

\bibitem[{{S{\'a}nchez-Bl{\'a}zquez}
  {et~al.}(2006{\natexlab{a}}){S{\'a}nchez-Bl{\'a}zquez}, {Gorgas}, {Cardiel},
  \& {Gonz{\'a}lez}}]{SB06a}
{S{\'a}nchez-Bl{\'a}zquez}, P., {Gorgas}, J., {Cardiel}, N., \& {Gonz{\'a}lez},
  J.~J. 2006{\natexlab{a}}, \aap, 457, 787

\bibitem[{{S{\'a}nchez-Bl{\'a}zquez}
  {et~al.}(2006{\natexlab{b}}){S{\'a}nchez-Bl{\'a}zquez}, {Gorgas}, {Cardiel},
  \& {Gonz{\'a}lez}}]{SB06b}
---. 2006{\natexlab{b}}, \aap, 457, 809

\bibitem[{{Sansom} \& {Northeast}(2008)}]{Sansom08}
{Sansom}, A.~E. \& {Northeast}, M.~S. 2008, \mnras, 387, 331

\bibitem[{{Schiavon}(2007)}]{Schiavon07}
{Schiavon}, R.~P. 2007, \apjs, 171, 146

\bibitem[{{Silva} {et~al.}(2008){Silva}, {Kuntschner}, \&
  {Lyubenova}}]{Fornax_red}
{Silva}, D.~R., {Kuntschner}, H., \& {Lyubenova}, M. 2008, \apj, 674, 194 (S08)

\bibitem[{{Thomas} {et~al.}(2003){Thomas}, {Maraston}, \& {Bender}}]{Thomas03}
{Thomas}, D., {Maraston}, C., \& {Bender}, R. 2003, \mnras, 339, 897

\bibitem[{{Thomas} {et~al.}(2005){Thomas}, {Maraston}, {Bender}, \& {Mendes de
  Oliveira}}]{Thomas05}
{Thomas}, D., {Maraston}, C., {Bender}, R., \& {Mendes de Oliveira}, C. 2005,
  \apj, 621, 673

\bibitem[{{Toloba} {et~al.}(2009){Toloba}, {S{\'a}nchez-Bl{\'a}zquez},
  {Gorgas}, \& {Gibson}}]{Toloba09}
{Toloba}, E., {S{\'a}nchez-Bl{\'a}zquez}, P., {Gorgas}, J., \& {Gibson}, B.~K.
  2009, \apjl, 691, L95

\bibitem[{{Trager} {et~al.}(2000){Trager}, {Faber}, {Worthey}, \&
  {Gonz{\'a}lez}}]{Trager00b}
{Trager}, S.~C., {Faber}, S.~M., {Worthey}, G., \& {Gonz{\'a}lez}, J.~J. 2000,
  \aj, 120, 165

\bibitem[{{Vazdekis} {et~al.}(2003){Vazdekis}, {Cenarro}, {Gorgas}, {Cardiel},
  \& {Peletier}}]{Vazdekis03}
{Vazdekis}, A., {Cenarro}, A.~J., {Gorgas}, J., {Cardiel}, N., \& {Peletier},
  R.~F. 2003, \mnras, 340, 1317

\bibitem[{{Wallace} \& {Hinkle}(1996)}]{WH96}
{Wallace}, L. \& {Hinkle}, K. 1996, \apjs, 107, 312

\bibitem[{{Worthey} {et~al.}(1992){Worthey}, {Faber}, \&
  {Gonzalez}}]{Worthey92}
{Worthey}, G., {Faber}, S.~M., \& {Gonzalez}, J.~J. 1992, \apj, 398, 69

\end{thebibliography}
\end{document}